\documentclass{PoS}
\pdfoutput=1
\usepackage[authoryear,square]{natbib}
\bibpunct{(}{)}{;}{a}{}{,}
\usepackage{aas_macros}

\title{HI tomographic imaging of the Cosmic Dawn and Epoch of Reionization with SKA}

\ShortTitle{Tomographic Imaging of CD/EoR with SKA}

\author{\speaker{Garrelt Mellema},$^1$ Le\'on Koopmans,$^2$, Hemant Shukla,$^{3}$, Kanan K. Datta,$^4$, Andrei Mesinger$^5$, Suman Majumdar$^1$ on behalf of
the CD/EoR Science Working Group\\
{$^a$}Dept. of Astronomy \& Oskar Klein Centre, Stockholm University, AlbaNova, SE-10691 Stockholm, Sweden;
{$^b$}Kapteyn Astronomical Institute, Universiity of Groningen, P.O. Box 9000, 9700 AA Groningen, The Netherlands;
{$^c$}Astronomy Centre, Department of Physics \& Astronomy, Pevensey II Building, 
University of Sussex, Falmer, Brighton BN1 9QH, UK; 
{$^d$}National Centre For Radio Astrophysics, Post Bag 3, Ganeshkhind, Pune 411 007, India; 
{$^e$}Scuola Normale Superiore, Piazza dei Cavalieri 7, I-56126 Pisa, Italy\\
        E-mail: \email{garrelt@astro.su.se}}

      \abstract{We provide an overview of 21cm tomography of the
        Cosmic Dawn and Epoch of Reionization as possible with
        SKA-Low. We show why tomography is essential for studying
        CD/EoR and present the scales which can be imaged at different
        frequencies for the different phases of SKA-Low. Next we
        discuss the different ways in which tomographic data can be
        analyzed. We end with an overview of science questions which
        can only be answered by tomography, ranging from the
        characterization of individual objects to understanding the
        global processes shaping the Universe during the CD/EoR.}

\FullConference{
Advancing Astrophysics with the Square Kilometre Array\\
June 8-13, 2014\\
Giardini Naxos, Italy}

\begin{document}

\section{Introduction}

SKA1-Low will provide enough sensitivity to deliver an image signal to noise 
larger than 1 for signals of order 1~mK on scales of a few arcminutes
for a combination of bandwidth and integration time $Bt\approx
1000$~MHz~hrs. This capability allows imaging of the redshifted 21cm signal
from the Epoch of Reionization on these angular scales. SKA1-Low will
be the first interferometer that will be able to do this as the
precursors LOFAR, MWA and PAPER have at least an order of magnitude
less sensitivity and thus can only hope to characterize the signals
statistically, for example through the value of the 21cm power spectrum at
certain scales or the variance of the signal at all observable scales
or higher-order statistics such as skewness and kurtosis.

This chapter which forms part of the set of chapters from the Cosmic
Dawn/Epoch of Reionization section of this volume discusses this
imaging capability of SKA1-Low and how it can be used to improve our
understanding of these two epochs. 

Actually as first discussed in some detail by
\cite{1997ApJ...475..429M}, since the signal can be imaged in many
adjacent frequency bins, it is better to speak of 21cm {\it
  tomography} when analysing the stacks of images, or image cube, at
different frequencies. The frequency direction is of course special as
along this line-of-sight direction the signal is seen from different
look back times (`light cone effect') and also is affected by redshift
space distortions introduced by the peculiar velocity field of the
gas. Still, the three-dimensional data will provide a measurement of
the distribution of the 21cm signal in three-dimensional space. In
this the 21cm signal is more akin to galaxy redshift surveys at very
much lower redshifts and very different from the Cosmic Microwave
Background which only provides us with a snapshot from a single epoch.

\begin{figure}[ht] 
\centering
\includegraphics[width=0.4\textwidth]{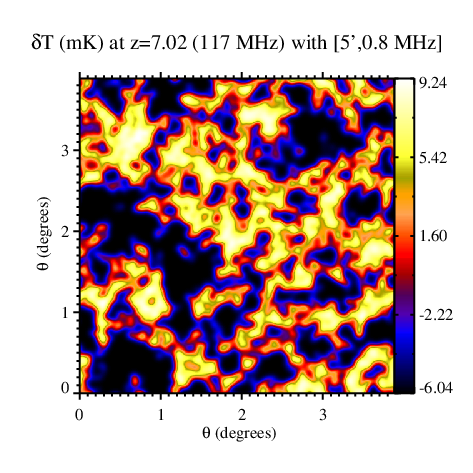}
\includegraphics[width=0.4\textwidth]{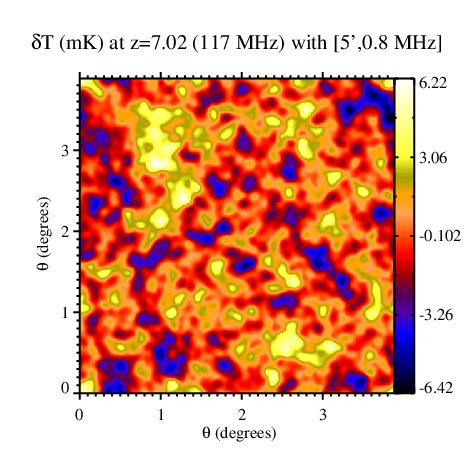}\\
\caption{\small Left panel: 21cm image at $z=7.02$ as derived from a
  full numerical simulation of reionization \citep[XL2
  from][]{2014MNRAS.439..725I}, convolved with a 5 arcminute FWHM
  Gaussian beam and 0.8 MHz bandwidth. The dark parts are large
  ionized regions. Right panel: a constructed 21cm signal with the
  same power spectrum as the signal in the left panel but a Gaussian
  PDF. These two images are very different but indistinguishable in a
  statistical power spectrum analysis.}
\label{fig:gaussian_signal}
\end{figure}

The 21cm signal also differs in another fundamental way from the CMB
signal. Whereas the CMB fluctuations reflect the tiny density
fluctuations around the time of decoupling, the 21cm signal has
strongly non-linear fluctuations. Furthermore, whereas the CMB
fluctuations follow a Gaussian distribution and thus are statistically
fully described by a power spectrum analysis, the same is not true for
the 21cm signal whose probability distribution function (PDF) has been
shown to be strongly non-Gaussian \citep{2006MNRAS.372..679M}. So,
although a power spectrum analysis can reveal important properties of
the signal, it does not fully describe it. This is illustrated in
Fig.~\ref{fig:gaussian_signal} which shows a 21cm image derived from a
full numerical simulation of reionization and an image with the same
power spectrum but a Gaussian PDF. The two images are very different but
at the same time are indistinguishable in a power spectrum analysis.

\section{The 21cm signal \& sources of variations}

As for example described in more detail in \cite{2006PhR...433..181F},
the redshifted 21cm signal from the intergalactic medium during the
CD/EoR is the differential brightness temperature $ \delta
T_\mathrm{b}$ which when scaled to canonical values can be written as
\begin{eqnarray}
  \delta T_\mathrm{b} & \approx &  27 x_\mathrm{HI} (1 + \delta)\left( \frac{1+z}{10} \right)^\frac{1}{2}
  \left( 1 -\frac{T_\mathrm{CMB}(z)}{T_\mathrm{s}} \right)
  \left(\frac{\Omega_\mathrm{b}}{0.044}\frac{h}{0.7}\right)
  \left(\frac{\Omega_\mathrm{m}}{0.27} \right)^\frac{1}{2} \nonumber\\
  & & \quad\quad\left(\frac{1-Y_\mathrm{p}}{1-0.248}\right)
  \left( 1 + \frac{1}{H(z)}\frac{\mathrm{d}v_\|}{\mathrm{d} r_\|}\right)^{-1}
    \quad\mathrm{mK}, \label{eq:dTb_scaled}
\end{eqnarray}
with $x_\mathrm{HI}$ the neutral hydrogen fraction,
$\delta=\rho/\langle\rho\rangle - 1$, the overdensity in the baryon
distribution, $T_\mathrm{s}$ the spin (or excitation) temperature of the
21cm transition, $T_\mathrm{CMB}(z)$ the Cosmic Microwave Background
(CMB) temperature, $z$ the redshift of the signal, $\Omega_\mathrm{m}$
and $\Omega_\mathrm{b}$, the total matter and baryon density in terms
of the critical density, $h$ the Hubble parameter in units of
100~km~s$^{-1}$~Mpc$^{-1}$, $Y_\mathrm{p}$ the primordial helium abundance
by mass, and ${\mathrm{d}v_\|}/{\mathrm{d} r_\|}$ the proper gradient
of the peculiar velocity along the line of sight. This last term
represents the effect of redshift space distortions.

From this expression one sees that the 21cm signal varies with
position due to variations in the matter overdensity
$\delta$, the hydrogen neutral fraction $x_{\mathrm{HI}}$, the spin
temperature $T_\mathrm{S}$ and line of sight velocity gradient
${\mathrm{d}v_\|}/{\mathrm{d} r_\|}$. This forms the basis of the
analysis of the 21cm signal be it statistically or tomographically.


\section{Regimes for Imaging}
\label{sec:regimes}
Imaging becomes possible once the signal to noise (S/N) for a certain size
of spatial/spectral resolution element becomes larger than 1. Since the instrument noise will decrease
when forming larger and larger resolution elements, even the first generation
experiments such as LOFAR can in principle produce images, although
with very poor resolution. This was worked out in detail in
\cite{2012MNRAS.425.2964Z} where it was shown that LOFAR could
produce images with a resolution of $\sim 20^\prime$, whereas power spectrum
analysis should be able to reach angular scales of $\sim 3^\prime$.

Since the sensitivity of SKA1-Low varies with frequency, imaging will
not be possible on the same scales at all frequencies. Specifically,
as the sensitivity drops rapidly below the critical frequency the
imaging capabilities for $\nu < 100$~MHz quickly deteriorate. In this
regime larger image resolution elements will need to be used to reach the same
noise levels. For those regimes in which imaging becomes unfeasible,
statistical analysis with power spectra should be used.

\begin{figure}[ht!] 
\centering
\includegraphics[width=0.5\textwidth]{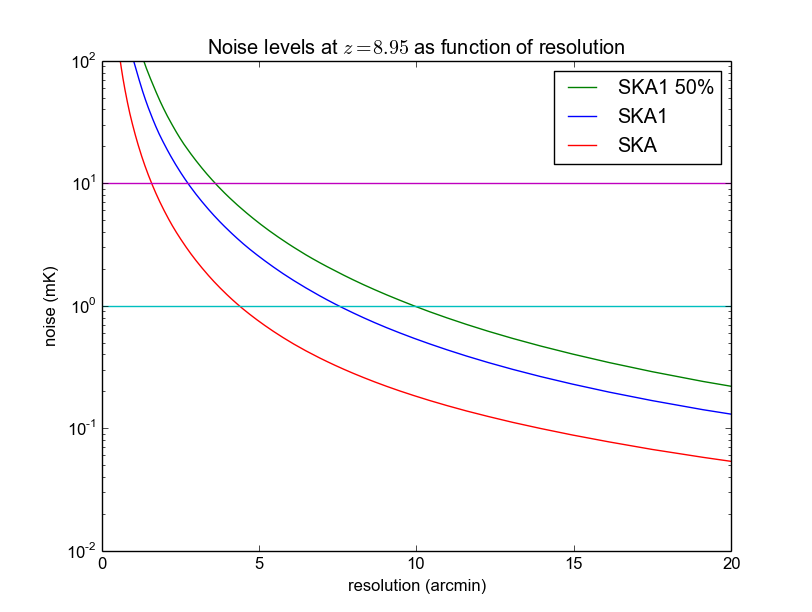}
\includegraphics[width=0.5\textwidth]{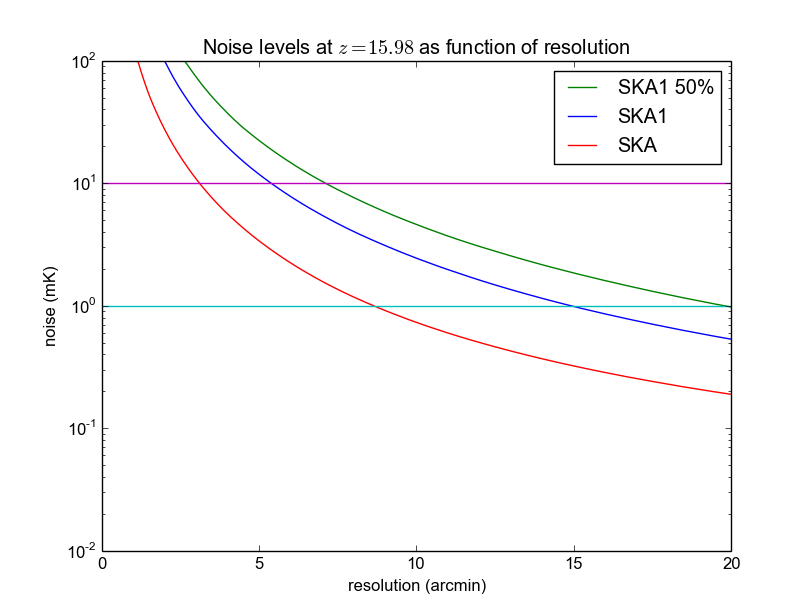}\\
\includegraphics[width=0.5\textwidth]{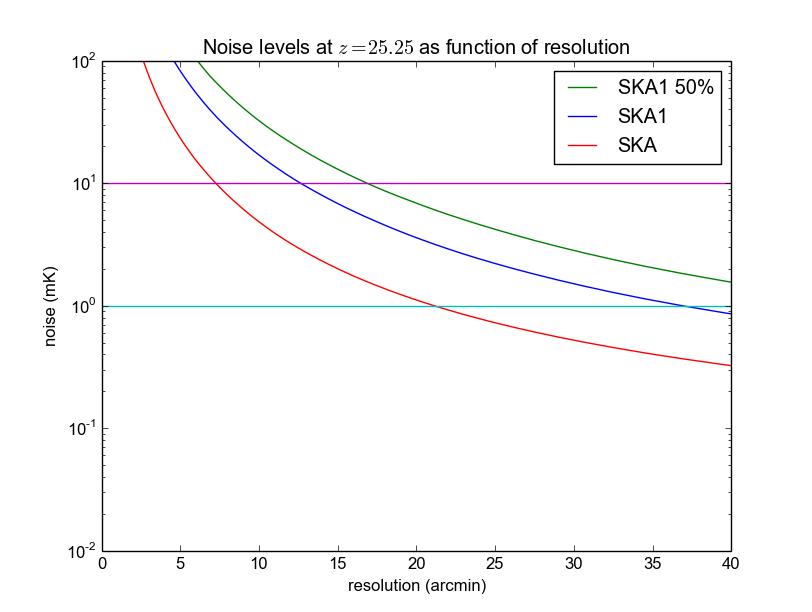}\\
\caption{\small SKA-Low image noise levels as a function of imaging
  resolution for redshifts 8.95, 15.98 and 25.25. The assumed
  integration time is 1000 hrs and the frequency bandwidth is matched
  to the angular resolution. The different curves indicate the early
  science SKA1-Low, the full SKA1-Low and the final SKA2-Low. The
  horizontal lines are placed at 1~mK and 10~mK noise levels. These
  results can also be found in \cite{Leon_Chapter}. The SKA1-Low
  configuration assumed is identical to the Baseline Design
  \citep{BLD_document}. Early science has half and the full SKA2-Low
  has four times the collecting area. }
\label{fig:tomography-SNR}
\end{figure}

The three panels of Figure~\ref{fig:tomography-SNR} illustrate the
imaging capabilities of SKA-Low in three different stages: an early
science 50\% of the Baseline Design, the Baseline Design SKA1-Low
\citep{BLD_document} and the full SKA2-Low at 4 times the sensitivity of
SKA1-Low. All cases are for an integration time of $1000$~hours and
for a frequency bandwidth matched to the angular resolution. For
tomography it is better if the frequency bandwidth matches the imaging
scale so as to have fairly uniform resolution across the
three-dimensional volume.

Concentrating on the full SKA1-Low case the figure shows that for $\nu
\sim 140$~MHz ($z\sim 9$) imaging on scales down to $7^\prime$
(19~comoving Mpc\footnote{For these conversions the cosmological
  parameters from Eq.~\ref{eq:dTb_scaled} are used.}) gives 1~mK noise
levels. For the regime around 80~MHz this increases to scales of
around $15^\prime$ (47~cMpc) and only very coarse imaging at scales of
0.5~degree (100~cMpc) or worse is feasable for the lowest frequencies.

At noise levels of 1~mK it is possible to detect fluctuations
in the neutral HI distribution. However, inspection
of Eq.~\ref{eq:dTb_scaled} shows that much larger amplitude variations
can occur in the signal. For example, for the case of $T_\mathrm{s}
\gg T_\mathrm{CMB}$ the contrast between fully ionized and fully
neutral regions at mean density is given by
\begin{equation}
   27\left( \frac{1+z}{10} \right)^\frac{1}{2} \quad\mathrm{mK}\, , 
   \label{eq:dTb_contrast}
\end{equation}
which should be detectable at higher noise levels than 1~mK.
It is thus possible to map out the shapes of ionized regions at higher
resolution. Figure~\ref{fig:tomography-SNR} shows that for 10~mK noise, SKA1-Low
could image ionized regions at $\sim 4^\prime$ (11~cMpc) at $z\sim 9$.

Similarly, variations in the spin temperature during the
Cosmic Dawn can result in more than 100~mK variations as can be seen
from evaluating 
\begin{equation}
   27 \left( 1 -\frac{T_\mathrm{CMB}(z)}{T_\mathrm{s}} \right)\left( \frac{1+z}{10} \right)^\frac{1}{2} \quad\mathrm{mK}. 
   \label{eq:dTb_spincontrast}
 \end{equation}
 for $T_\mathrm{s}$ in the range 5 to 200~K. As can be seen from
 Figure~\ref{fig:tomography-SNR} 10~mK noise levels permit imaging
 with SKA1-Low at $\sim 6^\prime$ (19~cMpc) scales around $z\sim 16$
 and at $\sim 14^\prime$ (47~cMpc) scales around $z\sim 25$.

For the other two configurations, the 50\% SKA1-Low case requires
roughly a 50\% worse resolution and the SKA2-Low case allows roughly a
factor 2 better resolution for the same noise levels, see
Fig.~\ref{fig:tomography-SNR}.

\begin{figure} [ht!]
  \centering \includegraphics [scale=0.4]{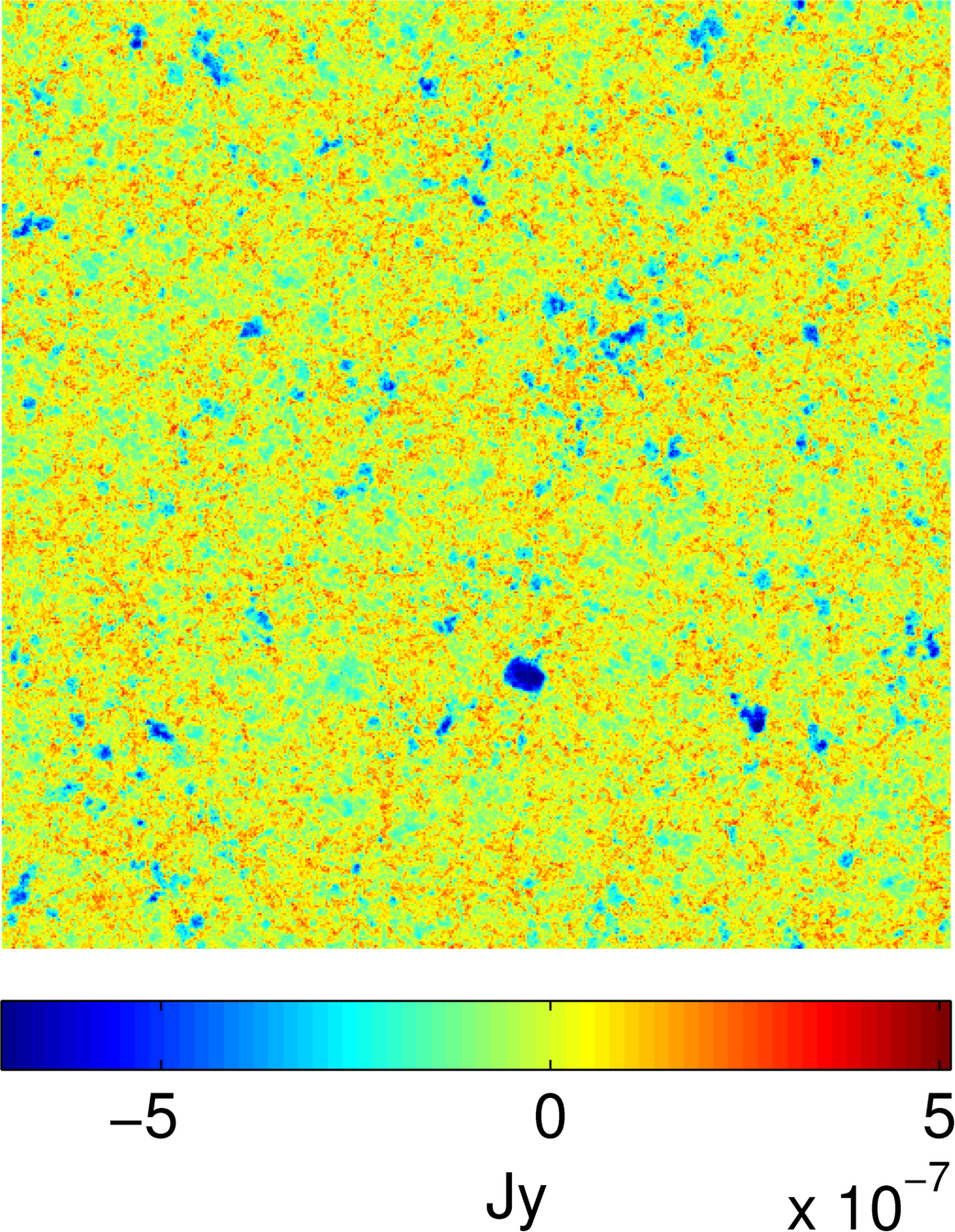}
  \qquad
  \centering \includegraphics [scale=0.4]{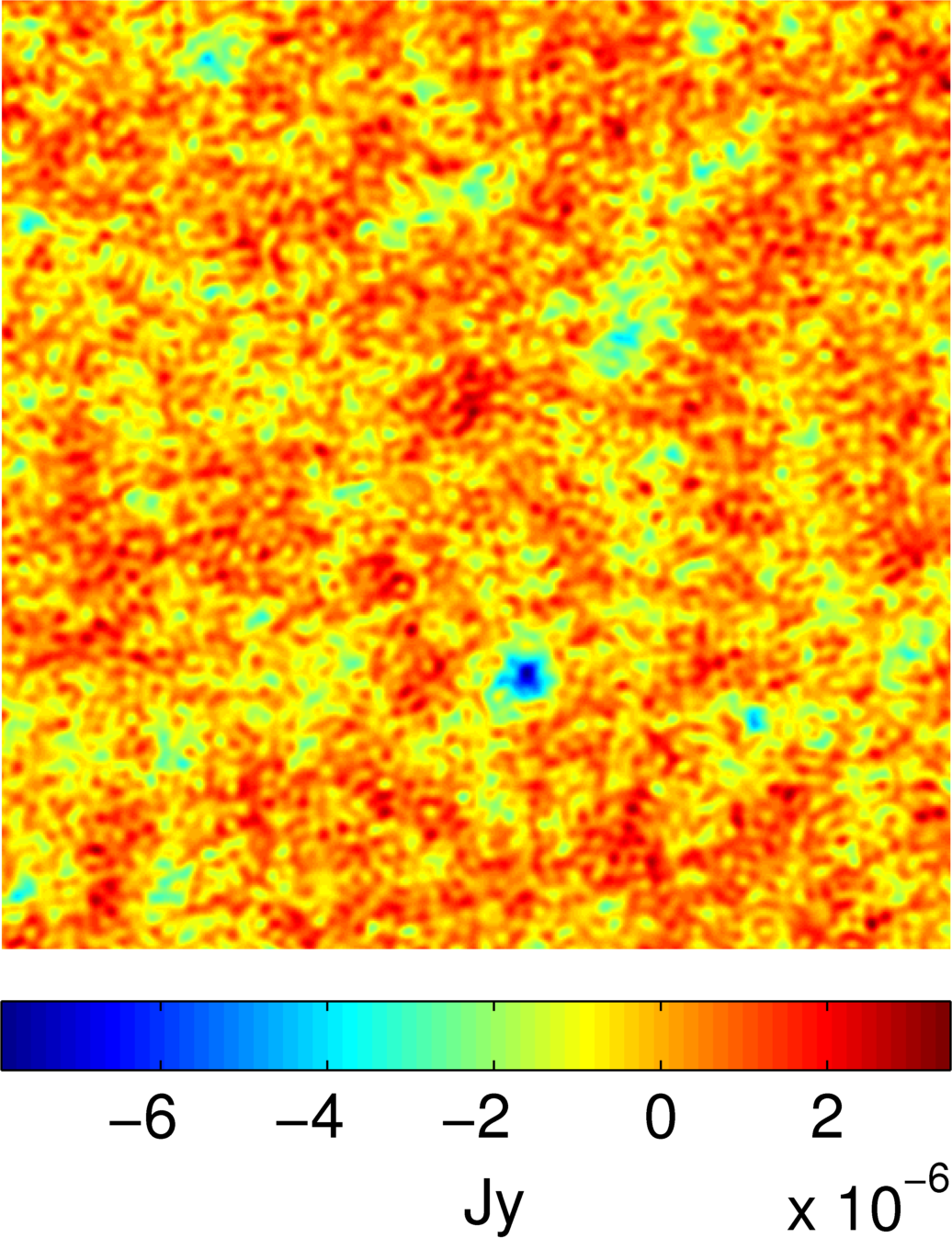}
  \caption[Caption for LOF]{Left panel: The 21cm signal from a
    simulation of reionization in a 425 $h^{-1}$~Mpc volume
    \citep{2014MNRAS.439..725I}. The redshift is $z = 8.515$ and it is
    assumed that $T_\mathrm{s}\gg T_\mathrm{CMB}$. The field is $3.64
    \times 3.64$ degrees on the sky. Right panel: the 21cm signal plus
    thermal noise when observed with SKA1-Low for 1000
    hours.\protect\footnotemark[2] The frequency is 150~Mhz, bandwidth
    1~MHz and only baselines less than 2~km are used. This gives a
    synthesized beam of $3.7^\prime \times 3.2^\prime$. The large HII
    region in the lower half of the image is $18.55$ $h^{-1}$~Mpc or
    $9.5^\prime$ in size and is clearly recovered in the observed
    image.}
  \label{Hemant}
\end{figure}

To further illustrate the imaging capabilities of SKA1-Low,
Figure~\ref{Hemant} shows the result of an imaging pipeline in which
an input image from a large scale reionization simulation has been
observed by an interferometer with the same specifications as in the
SKA1-Low Base Line Design \citep{BLD_document} for 1000 hours. This
pipeline is based on the MeqTrees software
\citep{2010A&A...524A..61N}. The largest HII regions in this image are
clearly visible (Shukla et al., in preparation).

\footnotetext[2]{For computational reasons only 10\% of the
    stations are used. To mimick the full SKA1-Low the noise is scaled
    down by a factor $10$.}\stepcounter{footnote}

\section{Analyzing images}
Once a tomographic data set has been obtained, it requires analysis to
draw quantitative conclusions about the (astro)physics of the CD/EoR.
Relying on the results of simulations several methods have been
considered for analyzing such three-dimensional data sets. However, as
the focus of the SKA precursors is on statistical measurements, the
analysis of tomographic data is still very much a field in
development.  In this section, we summarize some of the techniques
that can be used.  Section \ref{sec:science} provides an overview of
the science that can be done with tomographic data.


\subsection{Regions around special objects}
By targeting a region which contains a special object, such as a
bright quasar or a group of bright galaxies, the
21cm signal shows how this object has influenced its environment, either
through its ionizing radiation, or through its heating. The
measurement would provide the size and geometry of features in the
21cm signal (e.g.\ a spherical HII region around a bright
quasar). Such measurements would presumably employ rather ad hoc
analysis methods motivated by the interest in a specific area.

For the cases when the S/N is insufficient for imaging, for example
when the HII region is small or during the Cosmic Dawn when the noise
is high, a matched filter based method can be used to optimally detect
such regions using the visibilities directly, although this does require 
making assumptions about the shape of the region. This can then be used to
estimate their sizes \citep{2007MNRAS.382..809D, 2008MNRAS.391.1900D}.
Another technique that can be applied in the low S/N case is the
stacking of observations of several (similar) objects.

\subsection{Bubble size measures}
When not considering a special region, some sort of statistic is
needed to characterize the tomographic data set. For example, one
might want to characterize the distribution of HII bubble
sizes. Although this sounds quite straightforward, in reality this
will not be so. The reason for this is that the geometry of ionized
sources is unlikely to be a set of isolated nearly spherical bubbles
of various sizes. Simulations show that rather there exists a complex
network of ionized regions which partly overlap. The reason for this
is that the sources of reionization live in the developing
cosmic web and thus are spread out like beads on connected
strings. Simple two-dimensional cuts through this geometry do not do
justice to its complexity.

Several methods to characterize bubble sizes in simulation results
have been suggested. These include the ``spherical average method''
\citep{2007ApJ...654...12Z}, the power spectra of the ionized fraction
$P_{xx}$, the Friends-of-Friends (FoF) method
\citep{2006MNRAS.369.1625I} and PDFs (probability distribution
functions) of path lengths through ionized regions
\citep{2007ApJ...669..663M, 2014JKAS...47...49H}.  The results of the
first three methods were compared in \cite{2011MNRAS.413.1353F}. The
conclusion was that each method measures different aspects and
depending on what one is after, each provides a useful measurement of
bubble sizes.

It has to be noted that all of these methods rely on the ionized or
neutral fraction to characterize the bubble size distribution. The use
of these or similar methods on the redshifted 21cm signal from an
interferometer requires more investigation as the presence of density
fluctuations, noise, synthesized beam shape and possibly residual
foregrounds will affect their results.

\subsection{Topological measures}

A different property of the distribution of ionized regions is their
geometry and topology. These are often described through the so-called
Minowski Functionals which provide the total volume, total surface,
mean curvature and the so-called Euler Characteristic $\chi$. The
latter is a number characterizing the topology of the regions,
sometimes also expressed as the {\it genus} $g$, where $g=1-\chi$. If
$N_\mathrm{part}$ is the total number of isolated regions,
$N_\mathrm{tunnel}$ the number of tunnels through these regions and
$N_\mathrm{cavity}$ the number of cavities inside these regions, then
\begin{equation}
  \chi = N_\mathrm{part} - N_\mathrm{tunnel} + N_\mathrm{cavity}
\end{equation}
The work in \cite{2006MNRAS.370.1329G, 2008ApJ...675....8L,
  2011MNRAS.413.1353F, 2014JKAS...47...49H} suggests that such
topological measurements will be useful to characterize the
reionization process and distinguish between different
models. However, as for the bubble size methods discussed above, also
here more work is needed to adapt these methods for use with real 21cm
data. 

\subsection{Effects of finite resolution}
\label{sec:resolution}
Interferometers have both a minimum and maximum scale for which they
are sensitive and this will impact the images. Starting with the
minimum scale, finite resolution will limit the interpretation of the
data. Fluctuations in density, neutral fraction and spin temperature
below the resolution scale will not be separable. This means for
example that a resolution element with a low value of the 21cm signal can be either
a low density region or a high density region with several small,
unresolved HII regions. 

However, tomography does offer the possibility of analyzing this
further by creating an ionization mask. Large ionized regions should
be easily identifiable from their uniform, low flux level. Masking
these out from a statistical analysis, one can then derive power
spectra for the `neutral' regions. These regions will still contain
unresolved effects from small HII regions but comparing power spectra
from different redshifts would allow detecting their effects on the
density field. Note that such a masked analysis of the density field
is only possible with tomographic data supplying the location of the
masks.

As is well known, interferometers do not have baselines with zero
separation and thus are unable to measure the average value of the
signal within the FoV. In fact, there is a range of angular scales
that cannot be measured, determined by the length of the smallest
baseline. The smallest baseline for SKA1-Low has not yet been defined
but will presumably lie in the range of $\sim$30~m implying that modes
on scales of more than a few degrees will not be contained in the
tomographic data. It is therefore important that these missing scales do
not contain astrophysically or cosmologically essential structures as
it would be impossible to reliably reconstruct these from the
observations. Since we expect such information up to scales of a few
degrees, shortest baselines of $\sim$30~m are required for the CD/EoR
science case \citep{2013ExA....36..235M}. This implies station sizes
of at least this size, although the availability of correlations
between individual dipoles within stations would make it possible to
capture even larger scales.

\section{Science from images}
\label{sec:science}
In this section we discuss briefly some of the questions about
Reionization and the Cosmic Dawn that can be answered by analysing the
tomographic 21cm signal. Please also see \citet{Stuart_chapter} in
this volume for a case study of imaging with SKA1-Low.

\subsection{QSO properties}
By mapping out the shape of the ionized region around a bright quasar
several properties of the quasar can be derived. A requirement for
this is that the ionizing photon production from quasar dominates over
the that of the local galaxy population. As quasars presumably live in
biased, high density regions, this requirement is not trivially
fulfilled. However, for optimistic assumptions about the radiation
flux from a quasar, \cite{2012MNRAS.424..762D} do find that a quasar
can leave a clear imprint on its local environment. 


As they are expected to be large and prominent features in the
redshifted 21cm sky, the literature on quasar HII region in 21cm data
is extensive \citep{2005ApJ...634..715W, 2008MNRAS.386.1683G,
  2008MNRAS.387..469W, 2008ApJ...686...25F, 2012MNRAS.424..762D,
  2012MNRAS.426.3178M, 2013MNRAS.429.1554F}. The size of HII region
around the qusasar can be used to put constraints on the quasar life
time and luminosity, and possibly the mean neutral fraction of the
surrounding IGM. The occurence of fossil HII regions will put
constraints on the quasar population at high redshifts and will also
measure the properties of the ionized IGM.

A growing quasar HII region that is spherical in shape in the quasar's
rest frame, would appear to be distorted (`egg-shaped') along the line
of sight in a tomographic data set due to the light cone effect
\citep{1987ApJ...321L.107S, 2003AJ....126....1W, 2004ApJ...610..117W,
  2005ApJ...623..683Y}.  From this apparent anisotropy of its shape,
constraints on the quasar's ionizing luminosity, its age and neutral
fraction of the local IGM can be derived \citep{2008MNRAS.386.1683G,
  2012MNRAS.426.3178M}\footnote{We note that besides tomographic
  imaging the same can be achieved using an improved matched filter
  technique \citep{2012MNRAS.426.3178M} acting directly on the actual
  visibilities. This will possibly provide a better handle on the
  noise than the image cube \citep{2007MNRAS.382..809D,
    2013ApJ...767...68M} although the choice of the filter shape is
  important.} The constraints obtained from tomographic data will be
more stringent than the ones currently avialable from the quasar
absorption spectra \citep[e.g.][]{2011MNRAS.416L..70B}.

In addition such measurements will allow us to measure the possible
anisotropy of the emission from the quasar itself. If the ionizing
radiation only escapes in certain directions due to the presence of an
obscuring torus, as is seen in many lower redshift quasars, the
imprint on the surrounding medium will not be spherical. If such an
anisotropy would be found in the 21cm data, it would be the first
three-dimensional mapping of the radiative anisotropy of a high
redshift active galactic nucleus.

If quasars are present during the Cosmic Dawn, heating of the
surrounding IGM by their X-ray flux could leave strong imprints on the
21cm signal, creating 21cm emission regions against a background of
absorption. The contrast between the absorption and emission regions
could be as high as 200~mK which would make it possible to image these
even at quite high redshifts \citep{2010ApJ...723L..17A}, see also
section~\ref{sec:cosmicdawn} below.

\subsection{Connection between galaxy population and 
  HII regions}
Determining the volume and shape of the ionized region around a group
of optically detected galaxies would allow to infer the connection
between the ionizing radiation output of a galaxy population and the
brightest members of this population. It is well-known that the
galaxies we can currently detect are insufficient to reionize the
Universe \citep[see e.g.][]{2014ApJ...786...57S} and references
therein). However, the shape and extent of the luminosity function for
the unobserved population remains a matter of contention. By measuring
the specific examples of known galaxy groups, better constraints on
this problem can be derived. In addition such observations would map
the developing cosmic web.

As far as we know, no systematic exploration of such an approach has
been performed. Most of the work connecting galaxies and the 21cm
signal has focused on performing a cross-correlation between a galaxy
survey and a 21cm observation. This technique can even be employed in
noise regimes where imaging is not possible
\citep{2007ApJ...660.1030F, 2007MNRAS.375.1034W, 2009ApJ...690..252L,
  2013MNRAS.432.2615W, 2014MNRAS.438.2474P}.

\subsection{Bubble size distribution \& topology - breaking
  degeneracies in reionization models.}
Although power spectrum analysis will be an important tool to
determine the parameters of reionization, it can be expected that
there will be certain degeneracies between key parameters. As the 21cm
signal has a non-Gaussian distribution, power spectra do not fully
describe it and quite different models could be consistent with the
same power spectrum. It is likely although not proven that
tomographic data can break some of the degeneracies in a pure power
spectrum analysis. 

Both from the point of view of cosmology and galaxy evolution, the
important question is how to connect the measured size distribution
of HII regions to the dark matter halo distribution. This
correspondence will not be trivial as it involves a range of
(astro)physical processes, such as the feedback (thermal, mechanical
etc), recombinations etc. The measured bubble size distribution and
topology will shed light on all these mechanisms. However, to date a
detailed study of how to reliably extract this information from such
measurements remains to be done.

\subsection{Measurements of the cosmological density field}
As explained in Section~\ref{sec:resolution}, knowing the location of
large ionized regions would allow excluding these from the analysis of
the remaining signal, which would then be a modified version of the
cosmological density field. Since reionization is likely to occur
earlier in denser regions, the remaining signal would be dominated by
lower density regions and affected by unresolved ionized
regions. Still, a measurement such as this would provide useful
information, especially if complemented by higher redshift power
spectrum measurements. However, in order to produce high quality masks
and to detect significant density fluctuations, high resolution is
required. Likely this application will only become feasible with the
full SKA2-Low\footnote{We note that an alternative way to probe the
  large scale density distribution is to measure the dark matter power
  spectrum. It has been proposed that anisotropies in the HI 21 cm
  power spectra arising from the peculiar velocities (`redshift space
  distortions') can be used to extract the 'pure' dark matter power
  spectrum from the astrophysical HI power spectra
  \citep{2005ApJ...624L..65B, 2013PhRvL.110o1301S}.}.

\subsection{Measuring the global 21cm signal}
\label{sec:globalsignal}
Since the intrinsic signal from the ionized regions will be zero or at
least very close to zero, the presence of well resolved ionized
regions could be used as a zero-point to determine the power contained
in scales above the those set by the minimum baseline of the
interferometer, $> 3^\circ$ (at 150~MHz) for SKA-Low. We do not expect
substantial power in the spectrum of 21cm fluctuations beyond these
scales and consequently the value will be nearly equal to the mean
21cm signal in the Universe at that epoch. Measuring the flux from
well-resolved HII regions will thus yield an independent measurement
of the global 21cm signal. This technique does rely on the presence of
large enough ionized regions with mininal contribution from unresolved
neutral structures within them. A similar constant `zero-point' could
possibly also be found during the Cosmic Dawn if one assumes that the
regions with the lowest signal are the coldest ones whose temperature
can be derived from adiabatic cooling due to the expansion of the
Universe.

\subsection{Cosmic Dawn Science}
\label{sec:cosmicdawn}
Before $z\sim 12$, the spatial variations in the 21cm signal are
expected to be mostly due to variations in the spin temperature. As
explained in Section \ref{sec:regimes}, cold regions could generate
high amplitude signals ($> 100$~mK). Therefore higher image noise
levels of 10 -- 30~mK are acceptable, allowing for higher resolution
images at these frequencies. The spin temperature evolution is thought
to have two phases, an initial one during which UV photons from the
first stars will decouple the spin temperature from the CMB
temperature and couple it to the (cold) gas kinetic temperature, thus
making the 21 cm signal observable against the background CMB. This
phase is followed by a phase in which the cold IGM will be heated by
X-ray radiation, turning the 21cm signal from absorption into
emission. Both phases will likely be patchy and thus excellent targets
for tomography \citep[see e.g.][]{2014MNRAS.439.3262M}).

To start with the heating phase, the heating profile and the 21 cm
signal around X-ray sources will depend strongly on the flux and
spectrum of the X-ray radiation reaching the IGM. This in its turn
will depend on the intrinsic X-ray flux and spectrum and the level of
absorption within the host galaxies. For mini-QSOs or a hot
interstellar medium in star-forming galaxies, which emit substaintial
amounts of soft X-ray ($<1$ Kev) photons, heated regions will be
smaller and the heating will be patchy \citep{2014MNRAS.443..678P,
  2014arXiv1406.4157G}. On the other hand, heating will be more
homogeneous and heated regions will be very large if the spectrum is
dominated by hard X-ray photons, for example in the case of high mass
X-ray binaries \citep{2014Natur.506..197F} and/or a high level of
absorption within the host galaxies. Thus the measurement of sizes
of heated regions will be a direct probe of the first X-ray sources
and their spectra. This measurement will also probe the level of the
X-ray background \citep{2013JCAP...09..014C, 2014MNRAS.439.3262M}.

During the earlier decoupling phase the patchiness is driven by the
rarity and clustering of the sources of UV photons. Even though this
phase is expected around $z\sim 20$, the decoupled regions around
these sources may still be imaged with SKA-Low as the contrast
between coupled and decoupled regions will be high \citep[$\sim 100$~mK, see
for example][]{2007A&A...474..365S, 2008ApJ...689....1S,
  2014MNRAS.439.3262M, 2014arXiv1406.4157G}. Measurements of size and
distribution of the decoupled regions would probe the nature of the
very first stars and their environment.

One particularly interesting application of tomography for the early
Cosmic Dawn would be to map out the effects of `bulk flows', an
expected large scale variation of the velocity difference between dark
and baryonic matter \citep{2010PhRvD..82h3520T}. This effect would
modulate early star formation on scales of $\sim$100~Mpc,
corresponding to $30^\prime$ at $z=20$. Both the modulation in the UV
background and the heating could generate observable signatures
\citep{2012ApJ...760....3M}. In spite of the low sensitivity for imaging
around 65~MHz, such a modulation would actually fall within the
imaging possibilities of SKA-Low.


It should be pointed out that most of the papers cited above focus on
power spectrum analysis of the 21cm signal from the Cosmic Dawn. More
work is needed to establish the impact of different kinds of X-ray and
UV sources on the 21cm signal and how well these phenomena could be
imaged with SKA-Low.

\section{Conclusions}

SKA-Low will transform the study of the Cosmic Dawn and the Epoch of
Reionization by allowing tomographic imaging of the 21cm signal. With
this capability we will be able to address a range of questions not
accessible to a statistical, power spectrum characterization of the
signal. Deep observations of 1000 hours with SKA1-Low will provide
$\sim 1$~mK image noise levels at resolutions ranging from $\sim
7^\prime$ at $\sim 200$~MHz to $\sim 30^\prime$ at $\sim 50$~MHz,
assuming a frequency bandwidth matching in physical size. However, many
features such as isolates ionized regions during the EoR and isolated
heated regions during the CD, can be imaged at $\sim 10$~mK noise
levels, allowing image resolution elements in the range $4^\prime$ --
$14^\prime$ in the same frequency range.

Tomographic imaging will allow the characterization of individual
bright quasars, including their isotropy or lack thereof. It will also
make it possible to connect the 21cm observations to optical galaxy
surveys, thus unravelling the connection between photon producing
galaxies and the reionization process. The ionized bubble size
distribution and the topology of ionized regions uniquely characterize
the reionization process. Extracting these from the observations
requires tomographic imaging as they are not described by the power
spectrum of the 21cm signal, due to its inherent non-Gaussian
signature. Furthermore, identifying the location of ionized regions
will make it possible to study the baryonic density field during
reionization and using the ionized regions as a zero-point can be used
to determine the global 21cm signal. Tomographic imaging thus opens
the door to a wide range of studies inaccessible to a power spectrum 
analysis and essential to understand the reionization process.

The relatively coarse imaging possible for the highest redshifts may
still be able to characterize the sizes and distribution of 'heated'
and 'decoupled' regions and the effect of the velocity differences
between dark matter and baryonic matter during the Cosmic Dawn. This
will provide a unique and important probe of the very first X-ray and UV
sources. 

None of the existing low frequency precursors is capable of
tomographically imaging the redshifted 21cm signal and even the
planned HERA project will only attempt imaging at very large scales
and during the EoR. This makes SKA-Low in all its phases a truly
transformational telescope which will not only provide detailed 3D
tomographic imaging of the EoR but also coarse, yet groundbraking
imaging of the Cosmic Dawn. SKA-Low will be the only telescope on
Earth capable of looking this far back in the history of our Universe
and reveal the earliest phases of star and galaxy formation.

\bibliographystyle{apj}
\bibliography {imaging}


\end{document}